\documentclass[12pt,dvips]{article}

\usepackage{amsmath,amssymb,exscale}
\usepackage{array,multicol}
\usepackage{afterpage,float,flafter}
\usepackage{epsfig,rotating,pifont}
\usepackage{cite}
\@addtoreset{equation}{section} \makeatother

\title{
\vspace{1cm}
\Huge\textbf{Cosmic $D$-strings \\as Axionic $D$-term Strings}
\vspace*{.5cm}
\author{
\large \textbf{Jose J. Blanco-Pillado\footnote{email: blanco-pillado@physics.nyu.edu}, 
Gia Dvali\footnote{email: gd23@nyu.edu}~~and Michele Redi\footnote{email: redi@physics.nyu.edu}}\\
\emph{Center for Cosmology and Particle Physics,}\\\emph{Department of Physics, New York University}\\
\emph{4 Washington Place, New York, NY 10003}}}

\date{}
\begin{document}
\maketitle \thispagestyle{empty} \vspace*{.5cm}

\begin{abstract}

In this work we derive non-singular BPS string solutions from
an action that captures the essential features of a $D-$brane-anti-$D-$brane 
system compactified to four dimensions. The model we consider is a supersymmetric
abelian Higgs model with a $D-$term potential coupled to
an axion-dilaton multiplet. The strings in question are axionic $D-$term
strings which we identify with the $D-$strings of type II string theory. 
In this picture the Higgs field represents the open string tachyon of the $D-\bar{D}$ pair 
and the axion is dual to a Ramond Ramond form.  The
crucial term allowing the existence of non-singular BPS strings is the
Fayet-Iliopoulos term, which is related to the tensions of the $D-$string 
and of the parent branes. Despite the presence of
the axion, the strings are BPS and carry finite energy, due to
the fact that the space gets very slowly decompactified away from the core,
screening the long range axion field (or equivalently the theory approaches
an infinitely weak $4D$ coupling). Within our $4D$ effective action we also identify 
another class of BPS string solutions ($s-$strings) which have no ten dimensional 
analog, and can only exist after compactification.
\end{abstract}

\newpage
\renewcommand{\thepage}{\arabic{page}}
\setcounter{page}{1}

\section{Overview}

It was suggested long ago by Witten\cite{wittensuper} that fundamental 
superstrings (F-strings) of macroscopic length could
be observed in the form of cosmic strings. After the discovery of
$D$-branes\cite{joe},  it is natural to expect that a similar role
could be played by other extended objects of string theory such as
$D_{1+q}$-branes wrapped around internal $q$-cycles. It is interesting
that brane inflation\cite{braneinf} generically predicts
the formation of such objects, whereas the formation of point-like
or wall-like extended objects is suppressed\cite{tye1}.
Various aspects of the dynamics, formation and evolution of cosmic $F-$
and $D-$strings have been discussed in \cite{gia1, giaalex, cmp, gia2, damour, joe2}. Needless
to say that a possible observation of these objects would provide
a direct window into string theory.

Thus, both from an observational as well as from the fundamental
point of view it is important to understand the precise nature and
structure of the {\it stringy} cosmic strings.  This is the motivation
that led to the present work. We shall derive non-singular BPS string solutions which 
resemble many of the features of the $D-$strings. Following the conjecture 
in \cite{gia1}, which we review below, we interpret our solutions as the $D-$strings of 
string theory. Independently from the conjecture the solutions presented in 
this paper are new BPS objects which have interest of their own: they are the first example
of finite energy cosmic strings coupled to an axion field. 

In trying to understand the $4D$ picture of the $D-$strings, it is useful
to consider their description in terms of Sen's tachyonic vortices formed on
the worldvolume of a higher dimensional unstable $D-$brane-anti-$D-$brane 
($D\, - \, \bar{D}$) pair\cite{sen}. For example, a $D_1$-brane in ten dimensions can be
viewed as a tachyonic vortex on the worldvolume of a 
$D_3 \, - \bar{D}_3$ pair. This vortex  originates as follows. The gauge
symmetry of the system consists of two $U(1)$'s belonging to
the worldvolume theories of the $D_3$ and $\bar{D_3}$ respectively. The
tachyon ($\phi$), which is an excitation of an open string
stretched between the two branes, is charged under the diagonal
combination of the two $U(1)$'s.  When the branes annihilate, the tachyon condenses 
and this gauge symmetry is Higgsed. Since the tachyonic vacuum is topologically non-trivial \cite{ktheory}, 
there are topologically stable vortex configurations, analogous to Abrikosov-Nielsen-Olesen \cite{ANO} flux tubes, which carry magnetic flux of the Higgsed $U(1)$. These flux tubes are the $D_1-$strings.
In this picture we can understand their Ramond-Ramond (RR) charge 
as originating from the Wess-Zumino coupling
on the $D_3-\bar{D}_3$ worldvolume\footnote{In string theory descriptions the RR charge may also 
be reproduced by couplings of the tachyon to $C_2$.},
\begin{equation}
\int_{3+1} F_2\wedge C_2,
\label{wz1}
\end{equation}
where $C_2$ is the RR two-form and $F_2$ denotes the field strength of the diagonal $U(1)$
gauge field.  After compactification to four dimensions $D_1-$strings
become cosmic strings. Since in ten dimensions $D-$branes preserve half of the supersymmetries, 
in $4D$ it should be possible to find some kind of cosmic $D$-strings which are still 
BPS saturated objects.

It is therefore natural to ask whether there are such solitons in $4D$ supergravity.
In \cite{gia1} it was shown that the only BPS gauge strings in supergravity
are $D$-term strings. These are the strings that are formed by Higgsing a $U(1)$ gauge 
symmetry due to the presence of a Fayet-Iliopoulos (FI) $D$-term.  Because of this fact, 
it was conjectured that $D$-strings, if they have any solitonic counterpart
in the effective four dimensional field theory description, must be represented by some form 
of BPS $D$-term strings.

The "{\it $D$-string  $D$-term-string equivalence}" conjecture leads to the conclusion that the 
energy density stored in an unstable $D - \bar{D}$ pair is a $D$-term associated with
the FI term of the worldvolume $U(1)$ that is Higgsed by the tachyon.
Therefore, in the ten dimensional limit, the effective potential for the tachyon is schematically,
\begin{equation}
\label{Dterm1}
V_D \, = \, {g^2 \over 2} (\xi -|\phi|^2 +...)^2
\end{equation}
where the ellipses stand for all the other charged fields in the system.  These fields have positive
mass squared and vanish throughout the annihilation process. It follows from (\ref{Dterm1}) that 
the FI term $\xi$ is related to the $D_3$ brane tension by
\begin{equation}
\label{FIterm}
{{g^2\xi^2}\over 4} =  \, T_3 
\end{equation}
where $T_3$ is the $D_3$ tension and $g$ is the $U(1)$ gauge coupling constant.
In the ten dimensional limit, the annihilation proceeds solely through the tachyon condensation,
which compensates the FI term in (\ref{Dterm1}). The resulting $D-$strings are purely tachyonic vortices, 
carrying ten dimensional RR charge. Under the term {\it purely tachyonic} what is meant
here is that the only winding phase responsible for the topological stability of the string, and 
consequently for the existence of the magnetic flux, is the phase of the tachyon. From eq. (\ref{wz1}) 
it follows that the electric RR charge of the $D_1-$brane arises from the quantized magnetic flux carried
by the vortex.

The effect of compactification on the above system is rather profound.
First, after dimensional reduction we are left with the zero mode of the RR two-form, $C_2$, which in four
dimensions is dual to an axion field $a$. In a supersymmetric scenario the axion has a scalar partner 
$s$.\footnote{Depending on the details of the compactification $s$ is related to 
the dilaton and the volume modulus. In general the axion of the four dimensional theory arises from the zero mode of a
Ramond-Ramond field $C_{\mu \nu i...j}$ with two of the indices in the large dimensions.} Dualizing the 
two-form one generically obtains the following term in the action
\begin{equation}
\label{axiondual}
{{M_P^2}\over {s^2}} \left (\partial_{\mu} a \, + \, 2\, \delta\, A_{\mu} \right )^2,
\end{equation}
where $\delta\approx \xi/M_P^2$ and $M_P$ is the four dimensional Planck mass. From this we see that 
the Wess-Zumino term (\ref{wz1}) gauges the shift symmetry of the axion. 
From eq. (\ref{axiondual}) it is also clear that after compactification, the worldvolume $U(1)$ 
symmetry is always in the Higgs phase, even when the tachyon vacuum expectation value (VEV) is set to zero.    
The mass of the gauge field of course vanishes in the infinite space limit, as $\delta\to 0$.
Another effect of the compactification is that the $D$-term potential (\ref{Dterm1}) aquires
an $s$-dependent contribution,
\begin{equation}
\label{Dterm2}
V_D \, = \, {g^2 \over 2} \left(\xi \, - |\phi|^2 \,- {{\delta M_P^2}\over {s}}\right)^2 .
\end{equation}
Notice that this correction too dies off in the decompactified limit. This fact is consistent with the picture that
in ten dimensions $D$-strings are purely tachyonic vortices whose tension is set by the FI term $\xi$.

Previous attempts to derive solutions for $D$-strings as BPS solitons in four-dimensions 
were in the following directions. As we have already mentioned, in \cite{gia1} smooth BPS $D$-term string solutions 
were derived in the presence of a FI term $\xi$, but the dilaton-axion multiplet was not included in the solution.
Notice however that this is consistent with supersymmetrically removing this multiplet as  shown in \cite{gia2}. 
In \cite{deffayet,binetruy} axionic string solutions were studied in a model in which the
$U(1)$ anomalies are cancelled by Green-Schwarz (GS) mechanism. Although this model is primarily 
motivated by the heterotic $E_8\times E_8$ compactification, it nevertheless shares some similarities with the
type II string compactifications that are our main interest. The low energy effective actions are very similar,
with the crucial difference that in the heterotic theory the existence of a dilaton-independent
part of FI term $\xi$ is not obvious and was not included in \cite{deffayet, binetruy}. As a result, 
the string solutions found in these models are either singular or break all the supersymmetries. 

In this paper we show that the presence of the $\xi$ term in (\ref{Dterm2}) allows us to 
construct smooth BPS $D-$string solutions. Other than being new interesting BPS objects, 
their existence provides an additional support for the conjecture of \cite{gia1}. 
Some complementary string theoretic evidence for this conjecture was also provided 
in \cite{lawrence}, where $D-$term strings were identified in intersecting brane systems.
A discussion of $D-$term strings in the context of wrapped branes has also appeared in \cite{halyo}. 
Let us also mention that, while we only study in detail supersymmetric strings, the BPS properties of 
$D-$strings on a background such as flux compactifications \cite{cmp} or the deformed 
conifold \cite{klebanov} may be lost.
 
In the present work we focus on explicit field theoretic solutions
for $D-$strings and then interpret them in the framework of $D-$brane systems. Actually we find two
different types of smooth BPS cosmic string solutions, both of which have an interpretation in string theory:\\\\  
\vspace*{.4cm}
{\bf  $\phi-$strings ($D$-strings)}\\ The first type of solutions, which fits all the features of the $D$-strings,
is the tachyonic vortex. In this case the magnetic flux is induced by the winding of the tachyon phase,
and not by the axion. At large distances from the core the asymptotic values of the fields for this
solution are,
\begin{equation}
\label{fields1}
\phi \, \rightarrow \,  \xi e^{i n \theta}, ~~~~~~ s \rightarrow \infty,
\end{equation}
where $\theta$ is the angular coordinate in the plane perpendicular to the string.
As we will show the asymptotically infinite value of $s$ is the only way to avoid the logarithmic divergence
of the energy in this case. Such a divergence would be incompatible with the BPS condition. Since $s$ represents 
a combination of the dilaton and the volume modulus, the space gets decompactified (or the $4D$ effective coupling goes to zero)
at large distances from the core. This effect however happens very slowly since $s$ depends only logarithmically on the distance.\\\\
\vspace*{.4cm}
{\bf $s$-strings}\\ In the second  class of solutions the tachyon VEV
 vanishes at infinity, and the magnetic flux is induced by the winding of the axion.
$s$ now goes asymptotically to a constant in such a way to compensate the
$\xi$ term in the potential (\ref{Dterm2}).  These strings can be interpreted as the vortices produced
by Higgsing the $U(1)$ group of a $D-\bar{D}$ pair by compactification, as opposed to tachyon
condensation.  Thus, far away from the string core the tachyon mass is zero and the branes do not
annihilate. The tachyon may only develop an expectation value in the core of the string.
The existence of a BPS configuration on a background with a $D_3-\bar{D}_3$ pair may come as a surprise, 
because in ten-dimensions this background breaks all the supersymmetries (for an exception see \cite{karch}). 
However, the identification of the $D_3-\bar{D}_3$
tension with the $D$-term energy density suggests that tachyon condensation may not be the
only way for restoring part of the supersymmetries once the effects of the compactification are
taken into account.  Instead, the  FI term may be compensated by the dilaton $s$.
The corresponding BPS strings are then the $s-$strings.\\

\vspace*{.3cm}
This paper is organized as follows. In section 2 we describe the supersymmetric models that we will 
be considering. In section 3 we derive the Bogomol'nyi equations for the string configurations. In section 4 
we give a detailed description of the two types of cosmic string solutions in global as well as local 
supersymmetry. In section 5 some new features of the string solutions are discussed. In section 6 we 
interpret our solutions within the context of string theory. We end with some conclusions and future
directions.

\section{The model}
\label{model}

For the globally supersymmetric case the effective $N=1$ lagrangian that we will 
consider takes the following form in superspace (we follow the conventions in \cite{wb}),
\begin{equation}
L=\int d^4\theta \left[\Phi_i^\dagger e^{-q_i V} \Phi_i + K(S+\bar{S}+4
\delta V) +2 \xi V\right] +\int d^2\theta ~\frac 1 4 f(S) W^\alpha W_\alpha+ h.c.
\label{lagrangian}
\end{equation}
$\Phi_i$ are chiral fields with integer $U(1)$ charges $q_i$. Since the $U(1)$ symmetry 
is linearly realized on the $\Phi_i$ for simplicity we will consider the minimal K\"ahler potential 
for this superfield. The axion is contained in a chiral superfield $S$ whose lowest component 
is $s+ia$ where $s$ is (loosely speaking) the dilaton.
Invariance of the action under a constant shift of the axion demands that the K\"ahler potential $K$ is a function
of $S+\bar{S}$. This shift symmetry is gauged under the $U(1)$. 
In the string motivated models that we have in mind one typically finds K\"ahler potentials of the form 
$K=-M_P^2 \log\left[S+\bar{S}\right]$ (up to a numerical coefficient) 
so we will study this case in detail even though most of our results hold for any $K(S+\bar{S})$.

As far as the string theory motivation is concerned, we will use (\ref{lagrangian}) as the prototype 
lagrangian describing interactions of the open string tachyon $\phi$ (represented by lowest component 
of one of the $\Phi_i$) and the complexified RR axion $s + ia$ in type II 
compactifications with $D-\bar{D}$ background.   We shall show that these fields  are sufficient to describe 
non-singular BPS $D$-string solutions.  It is important to stress that (\ref{lagrangian}) is not the
complete low energy action, because a $D-\bar{D}$ system contains 
other light fields, e.g., such as the  brane-separation mode.  However, these additional fields 
can be consistently ignored.

Note that the action above is similar to the one of $E_8\times E_8$ heterotic string compactifications in which the 
$U(1)$ anomalies are cancelled by the Green-Schwarz mechanism, but with the important difference 
that we include a constant (moduli independent) FI term $\xi$.  For type II compactifications, such constant term
is necessary in order to account for the $D-\bar{D}$ tension, which should survive in the 
ten-dimensional limit. As we shall show this term is crucial for 
the existence of non-singular BPS $D$-string solutions. In this respect our action differs from the one considered in
\cite{deffayet,binetruy} in which no $\xi$ term was included, and consequently no smooth BPS 
string solutions were found. We do not include a superpotential, since BPS requirement 
demands that $F$-terms must be identically zero on the string solution\cite{gia1}.

In components the bosonic part of the action is,
\begin{eqnarray}
L&=&-|D_\mu \phi_i|^2- K_{S\bar{S}}|D_\mu S|^2-\frac
1 4 Re(f) F^{\mu\nu} F_{\mu\nu}+ \frac  1 4 Im(f) F^{\mu\nu}
\tilde{F}_{\mu\nu}-\frac 1 2 Re(f) D^2 \nonumber \\
D&=& -\frac 1 {Re(f)} \left(2\,\delta\,K_S- \sum_i{q_i}|\phi_i|^2+\xi\right)
\label{action}
\end{eqnarray}
with the covariant derivatives given by,
\begin{eqnarray}
D_\mu \phi_i&=&(\partial_\mu - i q_i A_\mu)\phi_i\nonumber \\
D_\mu S&=& \partial_\mu S+2 i \delta A_\mu.
\label{covariant}
\end{eqnarray}
Under a gauge transformation the bosonic fields transform as,
\begin{eqnarray}
\phi_i &\to& e^{i q_i \lambda} \phi_i\nonumber \\
A_\mu &\to& A_\mu+\partial_\mu \lambda\nonumber\\
a&\to& a-2 \delta \lambda.
\end{eqnarray}
From the last equation we deduce that the periodicity of the axion 
is,
\begin{equation}
\left[a\right]=4\pi \delta.
\label{periodicity}
\end{equation}

Depending on the charges $q_i$ we can distinguish two cases. When
the $\sum_i q_i=0$ the matter sector of the lagrangian is anomaly free. 
Gauge invariance of the action then fixes $f(S)$ to be a constant and
$\delta$ to be arbitrary. When the sum of the charges is different
from zero the theory has mixed anomalies. In any consistent theory all the gauge anomalies must vanish.
In this case the anomaly can be cancelled by the Green-Schwarz mechanism applied
to four dimensions \cite{witten} by choosing the gauge kinetic function $f(S)=S$.
The axion has a universal coupling of the form $a\,F\tilde{F}$. Under the gauge transformation
the axion shifts and generates a contribution to the anomaly which cancels
the one of the matter sector. Clearly this implies severe restrictions on the
representations which are automatically satisfied in string compactifications.
In heterotic string compactification one finds for example\footnote{We should note here that 
we are not aware of a mechanism to generate a $s$ independent FI term in heterotic string theory.
On the other hand in type II theories the FI term should account for the energy density of the $D-\bar{D}$ pair.},
\begin{equation}
\delta= \frac 1 {192 \pi^2}\sum_{i=1}^n q_i.
\label{delta}
\end{equation}

For the main part of this paper we will focus on the scenario with constant
gauge kinetic function, $f(S)=1/g^2$ and $K=-M_P^2 \log[S+\bar{S}]$ which is relevant for type II 
compactifications.  The pseudo-anomalous case requires only minor modifications.
The vacuum manifold corresponds to the zeros of the $D-$term in (\ref{action}),
\begin{equation}
M_P^2\frac {\delta} s +\sum_{i=1}^n q_i  |\phi_i|^2=\xi
\label{vacuum}
\end{equation}
Due to the fact that the axion shifts under the $U(1)$  gauge transformation
the theory is always in the Higgs phase in any vacuum.
In general the gauge boson eats a linear combination of the axion
and of the phase of the Higgs field while the orthogonal combinations remain massless.
The mass of the vector boson is given by,
\begin{equation}
m^2=g^2\left(\frac {4 \delta^2}{s^2}+\sum_{i=1}^nq_i^2 |\phi_i|^2\right)
\end{equation}
with $s$ and $\phi_i$ lying in the vacuum manifold (\ref{vacuum}).
In the supersymmetric vacuum the gauge boson belongs to a massive $N=1$
multiplet. 

\section{Bogomol'nyi equations}
\label{bpsstrings}

We consider solitonic string solutions with the fields only
varying in the transverse plane $(x,y)$. Remarkably for \emph{any} K\"ahler
potential of the form $K(S+\bar{S})$ the energy density of this configuration
can be organized in the Bogomol'nyi form,
\begin{eqnarray}
E&=&\int d^2x  |(D_x \pm i D_y) \phi_i|^2+ K_{S\bar{S}} |(D_x\pm i
D_y) S|^2+\frac 1 2 Re(f) (F_{xy}\pm D)^2\nonumber\\
&\pm&\xi\int d^2x F_{xy}.
\label{bogoenergy}
\end{eqnarray}
where we have integrated by parts assuming that the boundary term at infinity
is zero and is understood that the signs are chosen so that the energy is positive. 
From this we can directly read off the Bogomol'nyi equations,
\begin{equation}
(D_x \pm i D_y) \phi_i=0~~~~~~~~~~~~~(D_x\pm i
D_y)S=0~~~~~~~~~~~~~~~~~~~~F_{xy}\pm D=0
\label{bogo}
\end{equation}
where the covariant derivatives are given in (\ref{covariant}).
When the Bogomol'nyi equations are satisfied the tensions of these solutions
is minimal and proportional to the magnetic flux through the string,
\begin{equation}
T=\pm \xi \int d^2x F_{xy}.
\end{equation}
The minimization of the energy guarantees that the equations of motion are also satisfied.

Generically in a supersymmetric theory the Bogomol'nyi energy
bound signals that the solution leaves a fraction of the supersymmetries unbroken, i.e.
it is a BPS state. The same holds here. A configuration of the fields preserves some
supersymmetries if the variations of all the fermions is zero for
some projection of the supersymmetry parameters. The transformations
of the fermions under supersymmetry evaluated on the bosonic background
are given by,
\begin{eqnarray}
\delta_\epsilon \psi_i&=&i \sqrt{2} \sigma^\mu \bar{\epsilon}
D_\mu \phi_i\nonumber \\
\delta_\epsilon \chi &=&i \sqrt{2} \sigma^\mu \bar{\epsilon} D_\mu
S \nonumber \\
\delta_\epsilon \lambda &=& \sigma^{\mu\nu} \epsilon F_{\mu\nu}+i
\epsilon D
\end{eqnarray}
where $\psi_i$ and $\chi$ are the fermionic partners of the scalar fields and $\lambda$
is the gaugino.  It is straightforward to check that, when the Bogomol'nyi equations (\ref{bogo})
are satisfied, the solutions leave half of the supersymmetries unbroken, so as expected
these are BPS states. This guarantees their stability even quantum mechanically.
Supersymmetry immediately explains why the Bogomol'nyi equations
do not depend on the K\"ahler potential (except through the $D-$term), giving a rationale for why
it was possible to rearrange the energy in the simple Bogomol'nyi form (\ref{bogoenergy}).
It is also easy to see considering the supersymmetry variations, that in the presence of a
non zero $F-$term all the supersymmetries must  be broken. Therefore only $D-$term
strings can be BPS objects \cite{gia1}.
The only exception to this statement is when the superpotential generating the $F-$term is chosen such that
the system has $N=2$ supersymmetry because in this case there is no
distinction between $F$ and $D$ terms \cite{urrestilla}. However this mechanism does
not seem viable in supergravity because it appears to be impossible to construct 
supersymmetric actions with constant FI terms in $N=2$ supergravity 
with charged hypermultiplets \cite{kallosh}.

It is important to stress that when the BPS equations are satisfied
the tension is equal to the constant FI term $\xi$ times the magnetic flux.
This shows that any attempt to find smooth BPS configurations with a field dependent FI term is bound
to fail because this configuration would have zero energy and therefore it can only be the vacuum. 
Of course this leaves the possibility to build singular BPS string solutions.
 
\section{String solutions}

In this section we present the BPS string solutions of our model for the
relevant case $K=-M_P^2\log[S+\bar{S}]$. For simplicity we first consider the globally 
supersymmetric case (here we set $M_P=1$). We find two classes of string solutions, 
the ones created by the Higgs field $\phi$ and the ones created by the dilaton $s$ which have rather
different properties. We consider in some detail the case with constant gauge kinetic
function (corresponding to no anomalies). The case where anomalies are cancelled by the GS mechanism
is slightly more complicated but leads to qualitatively similar results that
are briefly reported in section \ref{pseudo}. Finally we show that our solutions 
can also be coupled to supergravity preserving their BPS nature.

\subsection{The $\phi-$strings  ($D-$strings)}
\label{phi}

These are the solutions where one of the fields $\phi_i$ goes to a finite
value at infinity. In order to solve the BPS equations for a string configuration
we take the following ansatz for the fields\footnote{The ansatz for the axion is chosen
to respect the periodicity obtained in (\ref{periodicity}).},
\begin{eqnarray}
\phi_1&=& f(r) e^{i n \theta}\nonumber \\
S&=&s(r)-2 i \delta m \theta \nonumber \\
A_\theta&=&n \frac {v(r)} r\,.
\label{ansatz}
\end{eqnarray}
We assume for simplicity that all the scalar fields except $\phi_1$
are equal to zero. One could construct more general BPS strings where several
$\phi_i$'s are different from zero. In fact our solutions are
reminiscent of semilocal strings \cite{semilocal} and the relation will be clarified in section
\ref{discussion}. Notice that, while it is consistent with the BPS equations (see below) to set
the other chiral fields to zero, $s$ cannot be a constant in a supersymmetric configuration
because in the BPS equations the dilaton is sourced by the gauge field.

Using the ansatz above for the fields the BPS equations read,
\begin{eqnarray}
f'&=&|n| \frac {1-q v} r f \nonumber \\
s'&=&-2 \,\delta\, \frac {|m| - |n|  v} r \nonumber \\
|n|\frac {v'} r&=&g^2 \left( \xi- \frac {\delta} s-q f^2 \right),
\label{bps}
\end{eqnarray}
where we have assumed $\delta$ and $\xi$ to be positive and we have set $f(S)=1/g^2$.
In general the first two equations imply the following relation between
$s$ and $f$,
\begin{equation}
s=-2 \delta \left(|m|-\frac {|n|} q\right) \log r-\frac {2 \delta} q \log f + k
\label{sf}
\end{equation}
where $k$ is an arbitrary integration constant related to the size of the core
of the defect. Taking into account the fact that close to the origin
$f\sim \alpha r^{|n|}$ (which follows from the first equation in (\ref{bps}))
and that $f$ goes to a constant at infinity one derives,
\begin{eqnarray}
s &\sim& -2 \,\delta |m| \log r ~~~~~~~~~~\it{as}~~~~~r \to 0\nonumber \\
s &\sim& -2 \,\delta \left(|m|- \frac {|n|} q\right) \log r~~~~~~~\it{as}~~~~~r\to \infty
\label{asymptotic}
\end{eqnarray}
At infinity $s$ always goes to infinity unless the winding of the axion
is $q$ times the one of the phase of the tachyon. In this case, 
there is family of solutions labeled by the asymptotic value of $f$. 
In all other cases the boundary conditions at infinity are $f(\infty)=\sqrt{\xi/q}$ and $v(\infty)=1/q$.

\begin{figure}
\label{fig1}
\centering\leavevmode \epsfysize=6cm \epsfbox{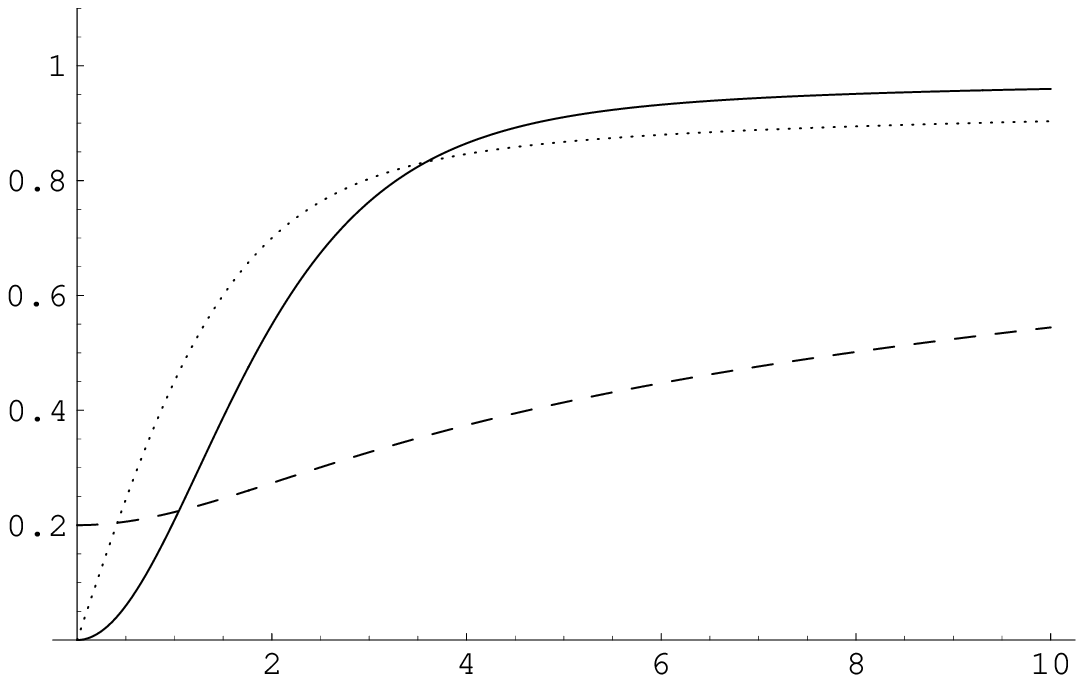}
\epsfysize=6cm    \epsfbox{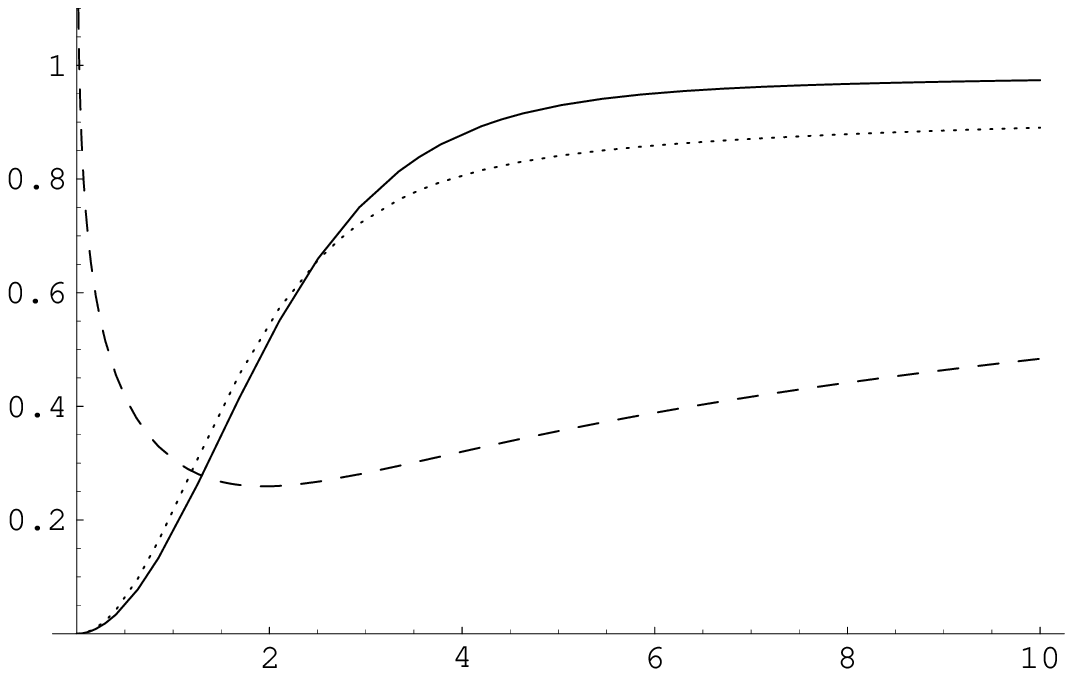}
\caption[Fig 1] {{\bf Plot of the $\phi$-string profiles.} {\it The figure on the left corresponds to the case $n=1$, $m=0$
described in the main text, while the one on the right corresponds to $n=2$, $m=1$. We have taken the following set of parameters 
to obtain these solutions numerically, $g=q=\xi=1$ and $\delta= 1/10$. The solid line denotes the function $v(r)$ that parametrizes the
magnetic field profile, the dashed line is the dilaton field $s(r)$ and the dotted line is the Higgs field $\phi(r)$.}}
\end{figure}

Restrictions on the possible values of $n$ and $m$ can be derived
from eqs.(\ref{asymptotic}).In any configuration with finite energy the dilation $s$ should never cross
zero. Also, $s\to 0$ corresponds to strong coupling in our effective theory
so we would not trust our solutions in that regime. Requiring that $s$ is always positive leads to,
\begin{equation}
q |m| \le |n|
\end{equation}
which fixes the charge $q$ to be positive. If $q$ is much bigger than one the only BPS
solution is with the axion not winding, $m=0$.
When $\delta$ is negative (the sign is determined a priori when the $U(1)$ symmetry
is anomalous) solutions can of course still be found but in this case we need to
choose a field with negative electric charge and $\xi$ to be negative. In a given
theory the string solutions can only exist if $\delta$ and $\xi$
have the same sign so in principle the coupling to the axion multiplet might
destroy the BPS solutions. In figure \ref{fig1} we plot the
string profiles for $n=1,~m=0$ and $n=2,~m=1$.

The asymptotic expansion for the Higgs and the gauge fields
can be derived by solving the linearized equations of motion
at infinity. To leading order one finds,
\begin{eqnarray}
f&=&\sqrt{\frac {\xi} {q}} -\frac {\sqrt{q}} {4\sqrt{\xi} (|n|-q |m|)}\frac 1 {\log r}+\dots \nonumber \\
v&=&\frac 1 q -\frac 1 {4 |n| \xi} \frac {1} {(|n|-q |m|)} \frac 1 {(\log r)^2}+\dots
\label{asymptotic2}
\end{eqnarray}
These expansions should be compared with the ordinary abelian Higgs model
where the fields approach their asymptotic values exponentially fast.
In comparison to the abelian Higgs model, where
the energy is localized an scales comparable to the inverse Higgs mass, here 
the energy is spread on larger scales (depending on $k$). Asymptotically our solutions
resemble axionic strings but the BPS equations guarantee that the energy
remains finite. The tension of the string is given by,
\begin{equation}
T=\pm \xi \int d^2x F_{xy}= \frac{2 \pi |n| \xi} q
\end{equation}
As this formula shows the tension depends only on the winding of the Higgs field
$\phi$ and not on the winding of $a$.

\subsection{The "s-strings"}
\label{s}

\begin{figure}
\centering\leavevmode \epsfysize=6cm \epsfbox{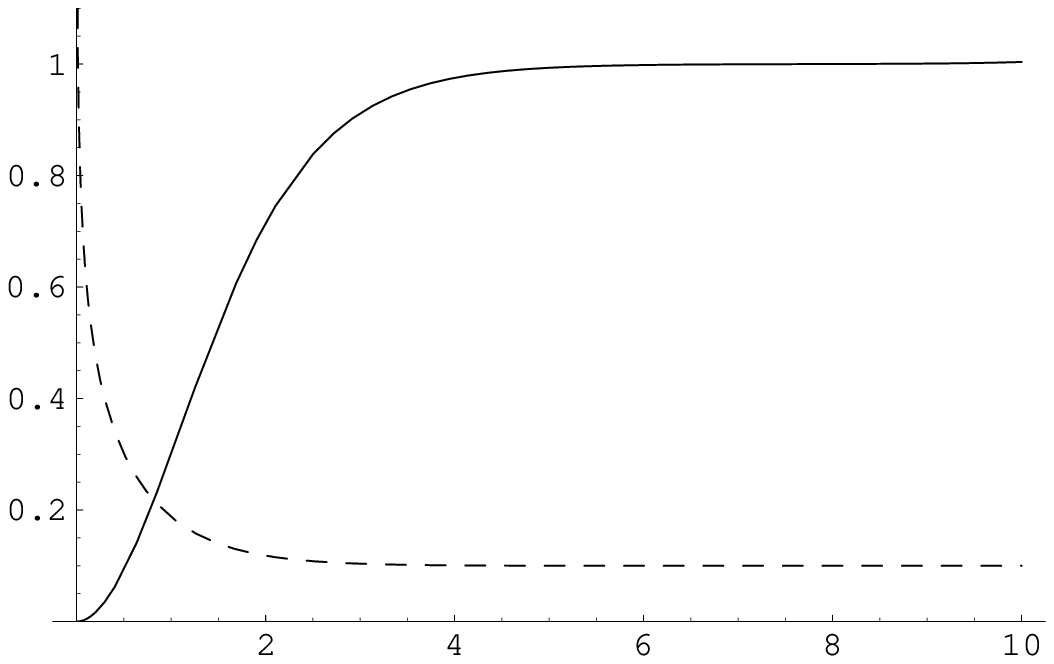}
\epsfysize=6cm \epsfbox{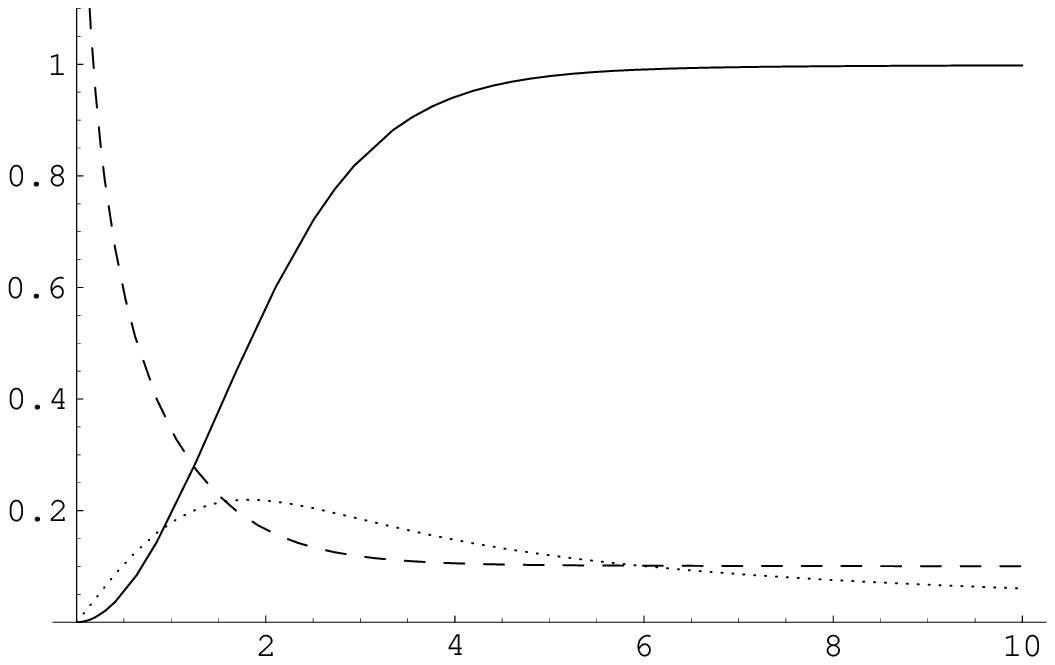}
\caption[Fig 2] {{\bf Plot of the $s$-string profiles.} {\it The figure on the left shows an $s-$string
with $\phi=0$ and $m=1$. The one on the right displays the case with $m=2$ and $n=1$. We use the same parameters
as in Fig.1.}}
\label{fig2}
\end{figure}

Very different solutions can be obtained when the string is supported by the field $s$ going
to a constant. In this case the ansatz for the fields is,
\begin{eqnarray}
\phi_1&=&f(r) e^{i n \theta}\nonumber \\
S&=&s(r)- 2 i \delta m \theta \nonumber \\
A_\theta&=& m \frac {v(r)} r\,.
\label{ansatz2}
\end{eqnarray}
This is almost identical to the previous case with the only difference
that the winding of the gauge field is now linked to the one of
the axion field. The BPS equations become,
\begin{eqnarray}
f'&=& \frac {|n|-q |m|v} r f \nonumber \\
s'&=&-2 \,\delta\, |m|\frac {1 - v} r \nonumber \\
|m|\frac {v'} r&=& g^2 \left(\xi-\frac {\delta} s- q f^2 \right)
\label{bps2}
\end{eqnarray}
From the first two equations it follows that, as in the previous case, $s$ and $f$ are related by
$s=-2 \delta(|m|-|n|/q) \log r-2 \delta/q \log f + k$.
This equation now requires that $f$ goes to zero at infinity since
$s$ goes to a finite value. The other boundary conditions which can be
deduced from the BPS equations are $v(\infty)=1$
and $s=\delta/\xi$. Numerical solutions are shown in Fig.(\ref{fig2}).

The tension of the $s-$strings is given by,
\begin{equation}
T=\pm \xi \int d^2x F_{xy}= 2 \pi |m| \xi\,.
\end{equation}
In general the tension of the $s-$strings is $q-$times the one of the $\phi$ strings.
The tension now depends only on the winding of the axion
and we have degenerate solutions with different values of $n$.
Let us consider the asymptotic expansion of these solutions. If $f=0$ the linearized
equations for $s$ and $v$ are identical to the ones for the abelian Higgs
model so the fields approach the asymptotic values exponentially. For 
$f\ne 0$ differently from the $\phi-$strings the fields approach
the asymptotic values power like,
\begin{eqnarray}
f&=&\frac {a} {r^{q |m|-|n|}}+\dots\nonumber \\
s&=&\frac  {\delta} {\xi}+\frac {\delta} {\xi^2}\frac {a^2} {r^{2 q |m|-2 |n|}}+\dots \nonumber \\
v&=&1 +\frac {|n|-q |m|} {|m| \xi^2} \frac {a^2} {r^{2 q |m|-2 |n|}}+\dots\,.
\label{asymptotic3}
\end{eqnarray}
As for the $\phi-$strings the relation between $s$ and $f$ leads to restrictions
on the values of $m$ and $n$, namely $q |m|> |n|$ (this simply follows from the fact that
since $s$ goes to a constant $f$ must go to zero).

In the special case $q |m|=|n|$, the $\phi$ and $s-$strings have the same tension.
In fact for these values of the parameters the two string solutions become degenerate
and it is possible to construct more general solutions with the fields approaching any
point of the vacuum manifold (\ref{vacuum}).

\subsection{Pseudo-anomalous $U(1)$ strings}
\label{pseudo}

Finally let us turn to the pseudo-anomalous case.
At least in some compactification the axion acquires an effective coupling,
\begin{equation}
\label{ffdual}
 a \, F_{2}^{A} \wedge \, F_{2}^{A},
\end{equation}
where $A$ runs over some gauge group factors $G_A$.
The existence of such a coupling signals that the fermion content of the theory 
has a mixed $U(1)\times G_A^2$ anomaly, which is cancelled by the Green-Schwarz mechanism. 

The coupling (\ref{ffdual}) has a double effect  on the $D$-strings solutions in question. 
First, it modifies the structure of the $D$-term potential, and secondly, because of instantons, 
it may lead to the attachment of the domain walls to the strings.  The $D$-term potential is 
modified, because in the case of GS anomaly cancellation $f(S)=S$.  
Using the same ansatz for the fields as in the previous cases the first two BPS equations remain unchanged
while the last equation becomes,
\begin{equation}
|n|\frac {v'} r=\frac 1 s \left(\xi-\frac {\delta} s-q f^2\right),
\end{equation}
where $n$ is the winding of the phase of the tachyon or of the axion
for the $\phi-$strings and for the $s-$strings respectively.

Let us first briefly discuss the $\phi-$strings. The vacuum
manifold is the same as in (\ref{vacuum}) plus the point $s=\infty$ (for
any value of $f$). $s$ diverges at infinity for the same reasons explained in
section \ref{phi}. This would in principle still allow $f$ to go to a constant
different from one at infinity. However this is incompatible with the equation above since
there are no solution where $v$ is bounded unless $f(\infty)=1$ (this can easily be
shown using the fact that $s\sim \log r$ asymptotically). The solutions of the BPS
equations are qualitatively very similar to the ones with constant
gauge kinetic function. In fact to leading order the asymptotic behavior of the
solutions is as in (\ref{asymptotic2}). For the anomalous $s-$strings the situation
appears even simpler. Since $s$ goes to a constant at infinity the solutions
are almost identical the ones in section \ref{s}.

Let us turn to the issue of the domain walls now.  The instanton effects may generate
a potential for the axion and the tachyon phase.  For instance, in the presence of 
the coupling (\ref{ffdual}), the instantons in question may be the gauge theory instantons  in one
or more of the $G_A$ groups.  The precise nature of the potential is model dependent, but the 
important fact is that the potential is only generated for the gauge-invariant combination of phases
\begin{equation}
\label{phasepot}
V(a\, + \, 2\, \Theta\delta)
\end{equation}
Thus, only the strings for which the combination $a\, + \,2 \Theta\delta$ winds at infinity can potentially become boundaries of the domain walls. 
For the heterotic $F$-strings the possibility of becoming boundaries of domain walls was suggested
by Witten\cite{wittensuper} and for the $D$-strings such a possibility was  discussed by 
Copeland, Myers and Polchinski\cite{cmp}. 
Presumably, the same physics that generates the potential (\ref{phasepot}) will also lift $\phi-s$ flat direction.
In such a case our $\phi-$ and $s$-string solutions  will acquire a logarithmically divergent energy, and will
become boundaries of the usual axion type domain walls.\footnote{One could also imagine a situation in which 
the $\phi-s$ flat direction is not lifted by the axion-potential. In such a case  the domain walls should  "dissolve" at infinity, 
because the potential should vanish in either of the two limits  $|\phi| \rightarrow 0$ and for $s \rightarrow \infty$. This follows 
from the fact that in both of these limits the physical gauge invariant axion becomes ill-defined. }

\subsection{Coupling to supergravity}
\label{supergravity}

In this section we briefly show how the string solutions presented in the 
previous sections can be coupled to supergravity.

Localized lumps of energy with codimension two create an asymptotically conical space-time with 
a deficit angle at infinity proportional to the tension of the object. At first sight this suggests that all
supersymmetries must be broken when gravity is dynamical because covariantly constant spinors 
cannot be defined on a conical space. This is the same problem faced and solved in the case of 
the pure abelian Higgs model  vortices \cite{becker,gia1} (see also \cite{edelstein} for related work).  
In the presence of a FI term the gravitino and the supersymmetric parameter $\epsilon$ become 
charged under the $U(1)$ gauge symmetry so that
the gravitino supersymmetric variation has an extra contribution proportional to the gauge connection which cancels exactly 
the one due to the deficit angle. In this section we generalize the analysis of the pure abelian Higgs model to a scenario with an
arbitrary K\"ahler potential and show that in general a BPS string configuration in global supersymmetry 
remains BPS when supergravity is (weakly) coupled.

We follow closely the analysis in \cite{becker} and refer the reader to
this paper for the details of the derivation. We take the following ansatz for the metric,
\begin{equation}
ds^2=-dt^2+dz^2+ e^{2\rho(x,y)} \left(dx^2 +dy^2\right).
\label{metric}
\end{equation}
In most general case \cite{wb} one can consider several chiral fields $A_i$
transforming under the $U(1)$ symmetry where the transformation corresponds to 
an isometry of the scalar manifold. In a bosonic background the
variations of the fermions read,
\begin{eqnarray}
\delta_\epsilon \chi_i&=&i \sqrt{2} \sigma^\mu \bar{\epsilon}
D_\mu A_i\nonumber \\
\delta_\epsilon \lambda &=& \sigma^{\mu\nu} \epsilon F_{\mu\nu}+i
\epsilon D\nonumber\\
\delta_\epsilon \psi_\mu&=&\left(\partial_\mu-\frac {i \xi}{2 M_P^2} A_\mu\right) \epsilon -\epsilon \omega_\mu -\frac 1 {4  M_P^2} J_\mu\epsilon
\label{gravitino}.
\end{eqnarray}
where $D_\mu A_j$ are the covariant derivatives of the scalar
fields in the theory which can be expressed in terms of the Killing vectors 
of the scalar manifold. For a general K\"ahler potential the current $J_\mu$ is given 
by,
\begin{equation}
J_\mu= K_j D_\mu A^j - K_{j^*} D_\mu A^{*j}.
\end{equation}
It is important to observe that since the gravitino is charged
under the $U(1)$ its supersymmetric variation contains a piece
proportional to the gauge connection. This is what allows to define
Killing spinors in an asymptotically conical space.

The supersymmetry transformations of the matter fields are as in
flat space with suitable factors of the metric. The BPS equations then become,
\begin{equation}
(D_x \pm i D_y) A_i=0~~~~~~~~~~~~~~~~~~~~~~~~~~~~e^{-2 \rho} F_{xy} \pm D=0.
\label{gravbps}
\end{equation}
Since the first equation does not depend on the metric explicitly,
for the model considered in this paper the same relation between
$s$ and $f$ found in (\ref{sf}) holds.

Supersymmetry is unbroken by the background if the Killing spinor
equations $\delta \psi_\mu=0$ admits solutions. These equations
imply the integrability condition \cite{becker},
\begin{equation}
-M_P^2\, \square \rho=\pm \xi F_{xy}+\frac i 2 \left(D_x J_{y}-
D_y J_x\right)
\label{ein00}
\end{equation}
where $\square$ is the two dimensional laplacian. This equation determines the conformal 
factor $\rho$, therefore what remains to be checked is that this equation is consistent with the
Einstein's equations for the metric. This is indeed the case as
one can check that (\ref{ein00}) is simply the $00$ Einstein equation for the
metric when the BPS equations (\ref{gravbps}) are imposed. The other equations
of motion are also satisfied as the energy can still be cast in the Bogomol'nyi form 
(\ref{bogoenergy}) (with appropriate metric factors). This concludes the proof 
that the BPS string solutions studied in this paper can be lifted to 
supergravity.

At large distance from the core of the string eq. (\ref{ein00}) can be easily
integrated since ($J_x, J_y$) go to zero faster than
$1/r$ (this assumption is necessary to write the  energy in the 
Bogomol'nyi form and is satisfied in our model). One finds that $e^{2\rho}=(x^2+y^2)^{-T/(2 \pi M_p^2)}$.
After a simple change of variables to polar coordinates the metric becomes,
\begin{equation}
ds^2_{r\to\infty}= -dt^2+dz^2+ dr^2+r^2\left(1-\frac T {2\pi M_P^2}\right)^2d\theta^2
\end{equation}
which as expected is the metric of a conical spacetime with
deficit angle $T/M_P^2$.

\section{Physical properties}
\label{discussion}

In this section we wish to describe some of the physical properties of the $\phi$ and
$s$ strings presented in the previous section. These are rather different from the ordinary abelian Higgs
model strings.

The large $r$ dependence of our solutions has a simple and important physical origin.
The contribution to the energy from the angular derivatives is given by ($M_P=1$),
\begin{equation}
\int r dr d\theta\frac 1 {r^2}\left[|\phi|^2(\partial_\theta\Theta-q A_\theta)^2+\frac 1
{s^2} (\partial_\theta a+2\delta A_\theta)^2\right]
\end{equation}
where $\Theta$ is the phase of the tachyon. With our choice of normalization the two phases
of the problem $\Theta$ and $a$ have periodicities equal to $2\pi$ and $4\pi \delta$ respectively.
It is then clear that for generic windings of $m$ and $n$ it is possible to cancel both
gradient terms at infinity only if we allow the factors $|\phi|^2$ or $1/s^2$ to go to zero. In fact this is the 
behavior of the fields required by the BPS equations but we would like to stress that the argument only relies
on the finite energy condition. 

We therefore have three different asymptotic strategies to find finite energy solutions. 
We can cancel the gradient energy associated with the tachyon field by choosing the gauge field 
appropriately and consequently have $s\to \infty$ as $r\to \infty$. These are the $\phi$-strings.
If the axion does not wind ($m=0$) the dilaton goes to a constant
at the core implying, contrary to the standard abelian Higgs model, that the gauge symmetry
is not restored there. The BPS equations however force the fields to be away from the vacuum
at the core. At infinity the gauge symmetry is Higgsed by the tachyon field instead.

On the other hand,  we can choose the gauge field such that it cancels the gradient 
term for the axion forcing $\phi\to 0$ as $r\to \infty$. These are the $s$-strings. 
These solutions resemble in some ways the stringy cosmic strings studied
in the early nineties \cite{vafa} which are also relevant for the description of the $D_7-$brane\cite{perry}.
In that case the only field is the complex dilaton $S=s+i a$ and finite energy supersymmetric configurations are
achieved by using the $SL(2,\mathbb{Z})$ invariance of the action. The solution for $S$ is 
a holomorphic function of the transverse coordinate $z=x+i y$. Close to the core,
\begin{equation}
S\approx -\log z
\end{equation}
which is the same asymptotic solution as for the $s-$strings. The divergence of the dilaton at the core was interpreted
in \cite{vafa} as decompactification of the space or weak coupling. As we will see in the next section a similar understanding will
be very useful to interpret our solutions as the $D-$strings of string theory.

The last possibility to achieve finite energy is to tune the winding numbers of the two fields 
in such a way that both gradients are canceled and therefore the string solutions can approach 
any point of the vacuum manifold at infinity.

The reader might have also noticed close similarities between the strings of this paper
and semilocal strings (see \cite{semilocal} and Refs. therein). In the
simplest semilocal strings there is a complex Higgs doublet with
equal $U(1)$ charges. String solutions can be found for example by setting one of the components of
the doublet to zero. In the Bogomol'nyi limit the strings can also be made supersymmetric.
In this case the stability of the strings is not guaranteed by the topology of the
vacuum manifold which is $S_3$ since $\pi_1(S_3)=I$. The same holds in our solutions.
Even though the vacuum manifold $M$ in our case appears different at first sight, we still have $\pi_1(M)=I$. The axion
phase becomes not well defined when $s\to \infty$ so the winding of the axion can be undone
by taking $s\to\infty$ without leaving the vacuum manifold. These solutions are stable because
of supersymmetry not because of topology. 

Another typical feature of the semilocal strings in the BPS limit, is the presence of  
zero modes \cite{leese} beside those due to the breaking of spacetime symmetries.
This is also present in our case since our solutions depend on an arbitrary integration constant
$k$ which controls the size of the string core. It is interesting to notice however that this does not lead to
extra localized zero modes at least for the $\phi$-strings.
To compute the effective action of the string we promote the
integration constant $k$ to a field depending on the coordinates
of the string. Considering the asymptotic expansion of the dilaton for the $\phi-$strings, $s\sim \log r +k$
it is easy to see that the kinetic term for $k$ is badly divergent (power-like). This
implies that this mode is frozen, it cannot be excited with any
finite energy excitation. It would be interesting to test numerically whether this mode 
can be excited in a collision of two strings or whether due to the power divergence 
of the kinetic term it is always frozen. For the $s-$strings the situation is different since
$f$ approaches the vacuum as a power law and therefore seems to be
possible (and model dependent) to have an integrable bosonic zero mode in this case.

Finally, we should also mention the presence of localized
fermionic zero modes. In the globally supersymmetric scenario there are at least 
two fermionic zero modes localized on the string due to partial
supersymmetry breaking. These modes are the supersymmetric 
partners of the translation modes living on the string.  Their profile can be 
obtained as usual by acting on the string background with the broken 
supersymmetry generators. In supergravity the story is rather different 
because, according to a well known argument
due to Witten \cite{wittensusy}, unbroken supersymmetry 
does not guarantee fermi-bose degeneracy. In fact it can be seen 
\cite{becker} that the fermionic zero modes become non-normalizable 
when gravity is dynamical so they are projected out of the spectrum
of the worldsheet theory.

\section{$D-$brane  interpretation}
\label{strings}

In this section we argue that the string solutions studied in this paper have a natural interpretation 
in terms of string theory solitons. To begin with let us review the key argument behind the conjecture of \cite{gia1}. 
According to this conjecture, in four dimensions any  non-singular BPS solitonic representation of $D$-strings (admitting an asymptotically 
locally flat space) must be some sort of a $D-$term string. This can already 
be argued in a globally supersymmetric limit,  but supergravity  gives the following general argument (see section \ref{supergravity}). 
Any finite tension (more precisely $T< M_P^2$) codimension two object, \emph{irrespective} of its core structure, 
creates a conical angular deficit at infinity. Covariantly constant spinors cannot be defined on a conical spacetime, so generically all the 
supersymmetries are broken. The only way to avoid this conclusion is if the gravitino is charged under the 
$U(1)$ responsible for the string magnetic flux, because in this case the gravitino variation
contains a piece proportional to gauge connection. The $U(1)$ in question is therefore a gauged 
$R-$symmetry. In $4D$ supergravity the necessary and sufficient condition
for the gravitino to carry a $U(1)$ charge is the presence of a constant FI term $\xi$.
This gives the gravitino a charge
\begin{equation}
q_{\psi}=\frac {\xi} {2 M_P^2}\,.
\end{equation}
Thus, if BPS $D$-strings admit a $4D$ solitonic description, they should be seen as $D-$term strings
with a FI term. This is the key point of \cite{gia1}. 

What is the physical origin of this FI term?  We have seen, that in theories with GS anomaly cancellation
there always is a dilaton-dependent  contribution to the FI term.  One may naively hope to convert 
this into a constant FI by stabilizing $s$. However, there is 
an obvious reason why such a program cannot work\cite{gia2}. Even if one imagines that 
there can be  a susy-preserving dynamics that stabilizes $s$, the same dynamics should also get rid of 
FI term at low energies.  This is because $s$ becomes part of a massive vector supermultiplet, 
and if we supersymetrically integrate out $s$, consistency demands that the whole vector field is integrated out
leaving no $D-$term at low energies.  Thus, the constant FI term responsible for the BPS $D$-strings must be there "from the beginning". 
The natural candidate is the tension of the $D -\bar{D}$ system. This conjecture passes several consistency checks, in particular 
it reproduces the correct relation between the $D-$brane tensions.   
However, the solution presented in \cite{gia1} was incomplete as BPS $D-$strings should also be charged under
Ramond-Ramond fields which were not included in that solution. 
Because of this, the above conjecture left some puzzling questions, which our solutions clarify.\\\\
\vspace*{.4cm}
{\bf $\phi$-String: Wrapped  $D_{1+q}$-Branes}\\
In ten dimensions $D_p-$branes are supersymmetric objects carrying a quantized charge of the
$C_{p+1}$ Ramond-Ramond gauge field. Upon compactification to four dimensions a $D_{1+q}$-brane 
can generate strings by wrapping a non trivial $q-$cycle of the internal manifold. These strings are 
will be charged under the two-form arising as the zero mode of the RR field to which the 
$D_{1+q}-$brane couples\footnote{In realistic compactifications, whether the  $D-$strings will remain charged under the
four dimensional RR two-form is model dependent as some zero modes may be projected out by 
the compactification \cite{louis}. Here we imagine a compactification
where the relevant zero modes are not projected out.}. In four dimensions a two-form is dual to 
a scalar so the $D-$string should be an axionic string. This naively leads to a puzzle because
if the string carries an axionic charge $\delta$, the energy density of the long range field should be
\begin{equation}
\frac {\delta} {r^2},
\end{equation}
which produces a logarithmic divergent energy per unit length.  While the formally infinite energy
of a single global string can be acceptable phenomenologically when several strings are included (such that the
total axion charge is equal to zero) the infinite contribution to the energy of the axion would certainly
destroy the BPS nature of the solutions.  This would seem to contradict to the expectation that $D$-strings 
are BPS saturated objects. 

The string solutions found in this paper have precisely the correct properties to solve this puzzle.
To be concrete let us consider $D_1$-strings produced in  $D_3-\bar{D}_3$ annihilation. 
When the two branes are far apart, that is at a distance greater than the string length $l_s$,
this system breaks all the supersymmetries. In the worldvolume description supersymmetry
is broken by the positive energy density of the branes. According to our description, 
this breaking  is reproduced in the low energy effective action by a constant FI term. 
The normalization is given by (in string frame)
\begin{equation}
2 T_3=\frac 2 {g_S (2\pi)^3  \alpha'^2}= \frac {g^2} 2 \xi^2,
\label{T3}
\end{equation}
where $g_S$ is the string coupling and $\alpha'=l_s^2$ (we follow the conventions of \cite{joebook}).
When the branes come closer than $l_s$ a complex tachyon develops in the spectrum of the
open strings stretching between the brane-anti-brane pair, signaling an instability of the system.
This is the origin of the tachyon $\phi$  in our effective action.
As explained earlier the gauge symmetry of the $D_3-\bar{D}_3$ pair is $U(1)\otimes U(1)$ and the tachyon is charged with respect
to the diagonal subgroup. The annihilation proceeds through tachyon condensation in which
the tachyon compensates the positive energy density restoring the supersymmetric vacuum without $D_3-$branes. 
This is the essence of Sen's tachyon condensation \cite{sen}. In the annihilation process codimension two objects 
are produced which are $D-$strings. 
In our picture the strings produced in the annihilation are simply the $\phi-$strings created by the winding of the phase of the tachyon.
One can see that the tension has the same scaling as in string theory. In our solution,
\begin{equation}
T_1=\frac 1 {g_S (2\pi) \alpha'}=\frac {2 \pi \xi} q\,.
\end{equation}
Comparing with eq. (\ref{T3}) this implies the relation $q^2 \, g^2=8\pi g_S$. This formula has the correct scaling 
between the Yang-Mills and string coupling. The numerical coefficient is also reproduced for $q=2$.

Let us now discuss the RR charges. The gauge field on the worldvolume of a 
$D_3$-brane couples to the RR two-form $C_2$ via the Wess-Zumino coupling,
\begin{equation}
2\pi \alpha' g_S T_3 \int_{3+1} F_2 \wedge C_2.
\end{equation} 
This coupling means that a magnetic flux tube carries the RR charge of $C_2$ in ten dimensions,
or in other words it is a $D_1-$brane.  In the $D_3-\bar{D}_3$ system branes have opposite RR charges 
so that $C_2$ couples precisely to the same linear combination under which the tachyon is charged. 
By compactifying to four dimensions only the zero mode of $C_2$ is kept, leading to an
effective action of the form,
\begin{equation}
\int d^4 x \left[M_P^2 (dC_2)^2+ \xi F_2\wedge C_2\right]
\label{action2}
\end{equation}
up to numerical factors and coupling to the dilaton.
In order to connect our solutions with the string theory picture it is useful
to dualize the axion $a$ to a two-form $C_2$ in the lagrangian (\ref{action}). 
The duality transformation is given by,
\begin{equation}
s^2 dC_2= \star (da+2\delta A)
\label{duality}
\end{equation}
through which the action takes the form (\ref{action2}).
One can deduce that the axionic charge under the $U(1)$ scales as 
$\delta\approx \xi/M_P^2$ which as expected vanishes in the large volume limit.

The four dimensional RR charge of the string solutions is given by,
\begin{equation}
\int \star dC_2
\end{equation}
where the integral is taken over a circle surrounding the string.
Using eq. (\ref{duality}) one can see that in all our solutions this integral is zero at infinity signifying that the
RR charge, as measured with respect to the zero mode of $C_2$, is always completely
screened at large distances. Notice however that the integral is not zero at finite distances
signaling the presence of an effective RR charge. 
This behavior, beside being forced by finite energy, has a natural interpretation in the full string theory.
After compactification the four dimensional field $s$ is a combination of the radius             
modulus and the ten dimensional dilaton. $s$ going to infinity therefore
corresponds to decompactification of the space (or alternatively the string coupling
going to zero). This is similar to the effect advocated in the stringy cosmic
strings \cite{vafa} which also plays a crucial r\"ole in the F-theory \cite{ftheory}.
Since the space is opening up new dimensions at large distances
it is not surprising that the RR charge goes to zero as measured with the zero mode of the
two-form $C_2$. 

Similar arguments can be used for the $D_{3+q}-\bar{D}_{3+q}$ system where $q$ dimensions are wrapped 
on  the internal manifold. In this case $D_{1+q}$ objects will be produced, which from
a four dimensional point of view are just $D-$strings. $D_{1+q}$ branes couple to a $C_{1+q}$ form.
Upon dimensional reduction the $D-$strings will couple to a two-form arising  as a zero mode
of $C_{1+q}$ so the analysis above can be repeated almost verbatim in this case.  

\vspace*{.4cm}
{\bf  The $s$-Strings:  $D-\bar{D}$ Bound State?} 
  
Let us now turn to the $s-$strings. The above discussion set the stage for a simple interpretation
of our $s$-string solution in the form of a BPS $D-\bar D$ bound state. In the $s$-strings the 
$D-\bar{D}$ energy is not compensated  by the tachyon, but by the dilaton field. 
Since the tachyon is not condensed in the vacuum, this suggests that the branes do not annihilate but rather
form some sort of boundstate. In ten dimensions the $D -\bar{D}$ system is unstable but this is not 
inconsistent with our findings since compactification is crucial in this case. In fact even in ten dimension 
a $D_2-\bar{D}_2$ pair can be made supersymmetric by switching on a constant electric field on the worldvolume \cite{karch}. In our case it is the magnetic flux that stabilizes the configuration so it would be interesting to investigate the 
relation between the two solutions. We should also mention that since asymptotically $s$ goes to a constant of order one, it might be that the effective theory breaks down due to strong coupling.\footnote{Note however that, since normally the existence of BPS objects visible at weak coupling can be extrapolated to strong coupling, it seems plausible
that these solutions have an interpretation in general.}

Even though we have focused on the $D_3-\bar{D_3}$ system we believe that our $s$-type string solutions are 
more generic and will exist in any compactification in which there is at least one surviving  $D_{3+q}$-brane 
with the Wess-Zumino type coupling 
\begin{equation}
\int_{3+q+1} \, F_{2} \wedge C_{2+q},
\end{equation}
between the worldvolume $U(1)$ field and the bulk RR $2+q$-form.  The topological reason 
for the existence of such strings is that in the effective $4D$ theory the $U(1)$ symmetry is Higgsed, 
provided the zero mode of $C_{2+q}$ form is not projected out. This topological argument was 
already noticed in \cite{tyegia} and \cite{cmp} but no smooth solutions were found.
From above it follows that the tachyon plays an irrelevant role  for the existence of the $s$-strings and 
its VEV can consistently be set to zero.  

\section{Conclusions and Outlook}
\label{conclusions}

The purpose of this paper was twofold. First, we tried to find explicit 
BPS-saturated string solutions in the supersymmetric abelian Higgs model coupled to an 
axion-dilaton multiplet. These are new BPS string solutions which arise in string 
motivated four dimensional models which have interest of their own.
Secondly, we gave evidence that the strings in question have the basic features to
describe the $D$-strings arising from unstable $D-\bar{D}$ systems in type II theories. 
In making this connection our guidelines were:  {\it 1)} Sen's picture 
in which a $D_{1+q}$ brane can be thought of as a vortex of the complex open string tachyon of a
$D_{3+q} -\bar {D}_{3+q}$ pair; {\it 2)}  
The conjecture \cite{gia1}  that in $4D$ the BPS $D$-strings are the $D-$term strings
formed by Higgsing the $U(1)$ gauge symmetry due to the presence of a constant FI term.    

One important result, supporting our string theory interpretation, 
is that finite energy BPS solutions can only
exist in the presence of a constant FI term in the effective
action. Any field dependent FI term in isolation, for example
those generated in heterotic string theory compactifications,
cannot support smooth BPS strings because these configuration would have
zero energy. The difference between the two cases arises from the
fact that a field dependent FI term is generated from the gauging
of the shift symmetry of the axion while the constant FI term
only affects the potential (in global supersymmetry). We have also
shown that the finiteness of energy forces the bosonic partner of the axion to vary
along the radial direction. This behavior has a simple physical
explanation. An infinite straight axionic string in four
dimensions is analogous to an electric charge in 2+1 dimensions so
it would naively have a logarithmically divergent energy per unit
length. The only way to avoid this conclusion is if the charge is
screened at large distances. This precisely achieved by the
variation of the dilaton. The charge as measured from the flux
goes to zero at large distances from the string so that the charge
becomes completely screened.

We have found two qualitatively different classes of BPS string
solutions which we have dubbed the $\phi-$strings and the
$s-$strings depending on whether the solution is supported by the
tachyon field $\phi$ or the dilaton $s$. For the $\phi-$strings
the magnetic flux tube of the string is induced by the winding of
the phase of the tachyon while for the $s-$strings what is
relevant is the axion winding. For special values of the windings
the two solutions merge into each other.

The coupling to supergravity was also discussed generalizing
previous studies of \cite{becker,gia1}. Under very general assumptions
we have shown that BPS string solutions in flat space (with any
K\"ahler potential for the chiral fields) remain supersymmetric
even after the coupling to supergravity. As usual a finite energy
configuration with codimension two produces an asymptotically
conical space with deficit angle proportional to the tension of
the object. The possibility to define Killing spinors in this
space relies on a remarkable cancellation between the deficit
angle and the Aharonov-Bohm phase as first noticed in 2+1 
dimensions in \cite{becker}. We have shown that this effect is
generic.

The main motivation for this work was the study of the connection
between $D$-term strings and stringy $D-$strings along the lines of \cite{gia1}.
$D-$strings carry RR charges so it was our goal to clarify how to couple 
the $D-$term strings to the RR fields. A string couples electrically to a two-form
which is dual to an axion in four dimensions. In this paper we
have shown that it is indeed possible to couple the abelian Higgs
model strings to an axionic multiplet preserving their BPS nature.

The relation between $D-$strings and gauge theory strings has interesting
consequences for string cosmology and phenomenology such
as the fact that brane-anti-brane energy is represented by an FI
$D-$term of the worldvolume theory. This observation relates $D-$term inflation \cite{dinf} to
$D-$brane inflation\cite{braneinf}. We hope to return to this and related issues in
a future publication. Also having explicit solutions could be useful for a better
understanding of the reconnection probability of cosmic $D$-strings and differentiating 
between ordinary gauge theory cosmic strings and the stringy ones\cite{joe3}. 
 
In this paper we heavily relied on supersymmetry and BPS properties for two main 
reasons. On one hand we wished to obtain exact solutions at least in some approximation. 
On the other hand, supersymmetry was an useful tool for "tagging" the solutions in the $4D$ 
theory and tracing their origin in full string theory picture. 
In the real world, of course, supersymmetry is broken, and this breaking is expected to affect 
some properties of our solutions.  For example, dilaton stabilization may create a divergent
tension. In this case the strings will acquire some properties of global axionic cosmic
strings, with an unusually small coupling to the axion.  Due to this smallness, the dynamics 
of such strings will still be dominated by their core structure. 

\section{Acknowledgments}

It is a pleasure to thank S. Chang, G. Gabadaze, A. Iglesias, R. Kallosh, J. Polchinski, A. Van Proeyen, 
N. Weiner for useful discussions on related topics. J.J. B-P. is supported by the James Arthur Fellowhip at NYU.
The work of G. D. and M. R. are supported in part  by David and Lucile  Packard Foundation Fellowship for  Science and Engineering, 
and by the NSF grant PHY-0245068.  

\vspace{0.5cm}   


\end{document}